\documentclass[twocolumn,showpacs,prl,amsmath,amssymb]{revtex4}
\usepackage[english]{babel}
\usepackage[T1]{fontenc}
\usepackage[cp1250]{inputenc}
\usepackage{graphicx}
\usepackage{dcolumn}
\usepackage{bm}
\usepackage{ulem}
\usepackage{amssymb,amsmath}
\newcommand{\vc}[1]{\mathbf{#1}}

\begin{document}

\title{Bound and free atoms diagnosed by the recoil-induced resonances:\\
1D optical lattice in a working MOT}

\author{Maria Brzozowska}
\author{Tomasz M. Brzozowski}
\author{Jerzy Zachorowski}
\author{Wojciech Gawlik}
\email{gawlik@uj.edu.pl} \affiliation{Marian Smoluchowski Institute of
Physics, Jagiellonian University, Reymonta 4, PL 30-059 Cracow}
\homepage{http://www.if.uj.edu.pl/ZF/qnog/}
\date{\today}

\begin{abstract}
We report on studies of simultaneous trapping of $^{85}$Rb atoms in
a magneto-optical trap (MOT) and 1D optical lattice. Using Raman
pump-probe spectroscopy we observe the coexistence of two atomic
fractions: the first, which consists of free, unbound atoms trapped
in a MOT and the second, localized in the micropotentials of the
optical lattice. We show that recoil-induced resonances allow not
only temperature determination of the atomic cloud but, together
with vibrational resonances, can also be used for real-time,
nondestructive studies of the lattice loading and of the dynamics of
systems comprising unbound and bound atomic fractions.
\end{abstract}


\maketitle

\section{I. Introduction}
Optical lattices~\cite{gryrob01} constitute a convenient tool for
studying phenomena related to periodically ordered systems. They are
formed by the light-interference pattern, characterized by spatial,
periodic light intensity and/or light polarization modulation. This
modulation leads to periodically varying dipole force which, acting
on the sufficiently cold atoms, can lead to their localization in
the minima or maxima of the light intensity distribution. Optical
lattices with spatially modulated light polarization provide energy
dissipation mechanisms resulting from cyclic optical pumping of the
traveling atoms between their ground state sublevels~\cite{dali89}.
This leads to atom temperatures below the Doppler cooling
limit~\cite{lett88}. Usually, optical lattices are filled with atoms
either prepared in a MOT and further cooled in optical molasses or
from the Bose-Einstein condensate~\cite{greiner01}. In principle,
lattice trapping can occur also in a MOT but this requires special
trapping beam configuration or complex phase stabilization schemes
\cite{petsas79, schad99}. Investigation of optical lattices can be
carried out by various methods. The most often used is the
pump-probe Raman spectroscopy~\cite{ver92,hemm93}. Other are based
on observation of fluorescence from lattice-captured
atoms~\cite{jessen92} and transient recoil-induced
 intensity modulation of the Raman probe
beam~\cite{kozuma95,kozuma96}.

However, when the lattice trapping occurs in less than three
dimensions or when the lattice potential is not sufficiently deep to
capture all atoms, atomic sample splits into two fractions, one
moving freely and the other one constrained, at least in one
direction. Guo and Berman suggested to use recoil-induced resonances
(RIRs) for studies of such systems~\cite{guo93} and analyzed
theoretically various specific pump-probe
configurations~\cite{guo93, guo92, guo94, guo95}. The first
experiments employed one- and two-dimensional lattices loaded from
optical molasses~\cite{ver92,hemm93}. In these experiments only the
atoms captured by the lattice were interrogated and no information
on unbound atoms was recorded. Still, as demonstrated in this paper,
the ratio between free and bound atoms can be used as a convenient
and sensitive indicator of the lattice dynamics and loading
efficiency.

Below, we describe the method we have developed for diagnostics of
such two-fraction systems and present results of our study. In our
configuration a 1D, blue-detuned, standing-wave optical lattice is
created directly in a regular working MOT, i.e. with trapping fields
constantly on. The configuration of 1D lattice we apply is the
simplest possible and imposes no polarization-gradient cooling. This
leaves magneto-optical trapping as the only temperature-controlling
mechanism which facilitates interpretation of the observations. We
have verified that, indeed, the lattice beams do not change the
atomic temperature, hence we can use them as Raman pumps for a
nondestructive, 2D RIR-thermometry~\cite{brzo052}. By means of the
Raman pump-probe spectroscopy we study the interplay between two
atomic fractions: one, which is too hot to be localized in the
lattice micropotentials, and the other, which is captured by the
lattice. To our knowledge this is the first observation of a robust
coexistence of an optical lattice with atoms trapped in a regular
operating MOT.

The paper is organized as follows. In Section II, we briefly recall
the principles of the Raman transitions between energy levels of a
two level atom, both for the case of a free, unbound particle as
well as for an atom captured in a lattice micropotential. In Section
III, we describe our experimental setup. In Section IV, the
experimental results and their interpretation are given. Section V
summarizes our work and indicates possible applications of our
method.

\section{II. Raman transitions in free and bound atoms}

\begin{figure}[h]
\includegraphics[scale=0.6]{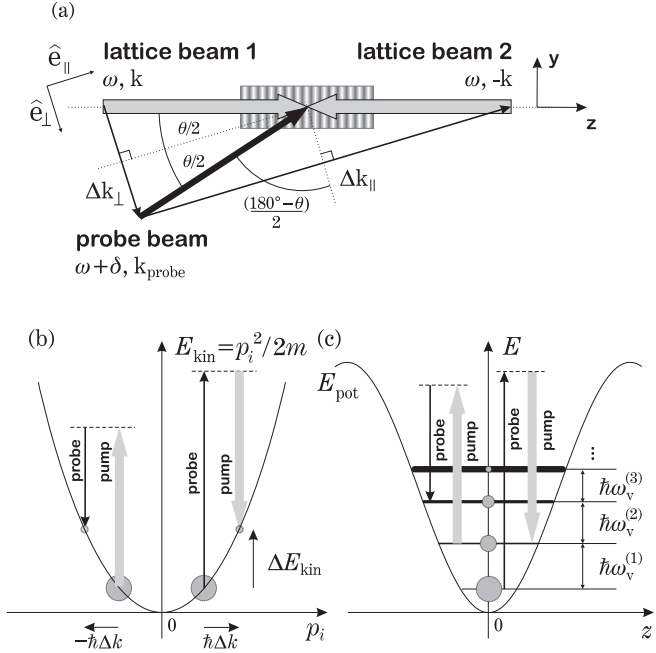}

\caption{\label{fig:raman} (a) Geometry of the light beams and
momentum transfers $\hbar\Delta\vc{k}_{||}$ and
$\hbar\Delta\vc{k}_{\perp}$ in the Raman transitions. (b) Raman
transitions between kinetic states of free atoms. (c) Raman
transitions between vibrational levels of atoms bound in the
optical lattice potential. The vibrational level spacing decreases
due to anharmonicity. In (b) and (c) circles symbolize populations
of the relevant levels.}
\end{figure}

We consider a two-level atom interacting with two counter-propagating laser
beams of frequency $\omega$ and of the same linear polarization, as depicted in
Fig~\ref{fig:raman}a. Since the beams produce a standing-wave interference
pattern, they act as lattice beams. A probe beam, of frequency $\omega+\delta$
and of the same polarization as the lattice beams, makes a small angle $\theta$
with one of the lattice beams. The probe beam detuning $\delta$ is scanned
around zero and the probe absorption is monitored.

The probe and lattice beams drive Raman transitions with absorption
of a photon from one beam followed by  stimulated emission to the
second beam (Fig.~\ref{fig:raman}b). Such processes couple two
kinetic states of a free-moving atom. As a result, the atom gains
kinetic energy $\Delta E_{\mathrm{kin}}=\hbar |\delta|$ and changes
momentum $\vc{p}$ by $\Delta \vc{p}=\hbar\Delta\vc{k}=\pm2\hbar k
\hat{\vc{e}}_i \sin\alpha/2$. Here, $k$ is the length of the light
wave vector, $\alpha$ is the angle between the probe and either of
the lattice beams: $\alpha=\theta$ for Raman transitions involving
the nearly co-propagating lattice beam, or
$\alpha=180^{\circ}-\theta$ for transitions involving the nearly
counter-propagating lattice beam, whereas $\hat{\vc{e}}_i$ is the
unit vector perpendicular to the bisector of $\alpha$. The sign in
$\Delta\vc{p}$ depends on the direction of the Raman process. The
resulting net variation of the probe beam intensity, $s(\delta)$,
depends on the population difference of the momentum states and
shows a resonant behavior around $\delta=0$~\cite{guo93, guo92,
guo94, guo95, gry94, ver96}. This recoil-induced resonance has a
shape of the derivative of the atomic kinetic momentum distribution
along the momentum exchange direction,
$s(\delta)\propto\partial\Pi(p_i)/\partial p_i$, where
$p_i=\vc{p}\cdot\hat{\vc{e}}_i$.

Since two lattice beams combine with the probe to drive the Raman
processes, two RIR signals are expected in the probe transmission
spectrum. The width of the RIR signal, proportional to the width of
the momentum distribution along relevant $\hat{\vc{e}}_i$, scales
with the angle $\alpha$ as $\sin(\alpha/2)$. Hence, assuming that
for given conditions the momentum distribution widths are of the
same order, we expect one narrow contribution, resulting from recoil
in the $\hat{\vc{e}}_{\perp}$ direction, and one wide contribution,
associated with recoil in the $\hat{\vc{e}}_{||}$ direction. The
presence of the two well-resolved recoil signals in the recorded
spectrum enables simultaneous determination of the momentum
distributions in the two orthogonal directions~\cite{brzo052}. When
$\theta$ tends to zero, $\hat{\vc{e}}_{||}$ ($\hat{\vc{e}}_{\perp}$)
becomes parallel (perpendicular) to the $z$ axis, which defines the
direction of the optical lattice confinement.

The above mechanism works well for atoms which move freely along
$\Delta\vc{k}$ and their kinetic energy can change in a continuous
way. In our case the motion of atoms confined in the 1D lattice
wells becomes quantized in the $z$ direction. This results in
vibrational structure of energy levels, unevenly spaced by
$\hbar\omega_{\rm{v}}^{(1)}$, $\hbar\omega_{\rm{v}}^{(2)}$,
$\ldots$, which differ because of anharmonicity of the potential
(Fig.~\ref{fig:raman}c). By averaging the vibrational level spacing
over anharmonicity with weights reflecting the temperature-dependent
populations of given vibrational levels one can find positions of
the most distinct Raman vibrational resonances~\cite{ver92,lett92}.
The first harmonic, $\delta=\pm\bar{\omega}_{\rm{v}}^{\,I}$, and
overtone, $\delta=\pm\bar{\omega}_{\rm{v}}^{\,II}$, are associated
with the change of vibrational quantum numbers by one and two,
respectively. Atoms trapped in the lattice do not give rise to the
wide RIR signal associated with movement in the $\hat{\vc{e}}_{||}$
direction. They participate only in the generation of narrow RIR and
Raman vibrational spectrum. Such a suppression of recoil in an
optical lattice, consistent with the previous
predictions~\cite{guo93, guo92, guo94, guo95} and
experiments~\cite{ver92,hemm93} suggests a novel method for
diagnostics of lattice loading by measuring the magnitude of the
appropriate RIR signal.

\begin{figure}[h]
\includegraphics[scale=0.8]{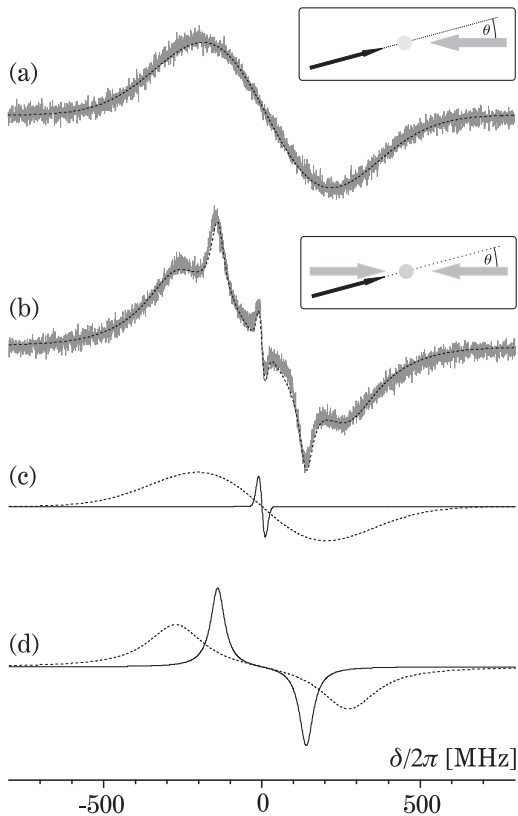}
\caption{\label{fig:spec_decomp}Experimental spectra recorded at
$T=64\,\,\mu\mathrm{K}$ and corresponding fits. (a) Single, wide RIR
spectrum (grey) acquired in the simple two-beam configuration shown
in the inset. Dotted black line represents the fit by the first term
of eq.\eqref{eq:fit}. Such spectra are used to determine temperature
of the atomic cloud. (b) Experimental spectrum (grey) recorded in
the presence of the 1D optical lattice (inset) and the fit by all
terms of eq.\eqref{eq:fit} (black). (c) RIR components of the fit:
wide RIR associated with the momentum exchange in the $||$ direction
(dotted line) and narrow RIR associated with the exchange in the
$\perp$ direction (solid line). (d) Vibrational components: first
harmonic (solid line) and overtone (dotted line), extracted from the
fit.}
\end{figure}

Various kinds of pump-probe spectra discussed in this Section are
illustrated in Figs.~\ref{fig:spec_decomp}a,~\ref{fig:spec_decomp}b
by the results of experiments described below in Section III. All
measured signals are fitted with the formula
\begin{align}\label{eq:fit}\nonumber
s(\delta)=&-A_{||}\frac{\partial
\mathcal{G}(\delta,\xi_{||})}{\partial
\delta}-A_{\perp}\frac{\partial
\mathcal{G}(\delta,\xi_{\perp})}{\partial
\delta}\\
&+L_I\left[\mathcal{L}(\delta,-\bar{\omega}_{\rm{v}}^{\,I},\gamma^{\,I})
-\mathcal{L}(\delta,\bar{\omega}_{\rm{v}}^{\,I},\gamma^{\,I})\right]\\
\nonumber
&+L_{II}\left[\mathcal{L}(\delta,-\bar{\omega}_{\rm{v}}^{\,II},\gamma^{\,II})
-\mathcal{L}(\delta,\bar{\omega}_{\rm{v}}^{\,II},\gamma^{\,II})\right],
\end{align}
where $\mathcal{G}(\delta,\xi)$ is the normalized Gaussian
describing kinetic momentum distribution of width $\xi$,  and
$\mathcal{L}(\delta, \omega, \gamma)$ is the Lorentzian used for
modeling the Raman vibrational resonance of width $\gamma$ centered
at $\omega$; $A_{||}$ and $A_{\perp}$ represent amplitudes of the
wide ($||$) and narrow ($\perp$) RIR contributions, and $L_{I}$ and
$L_{II}$ are the amplitudes of the vibrational resonances associated
with the first harmonic at $\bar{\omega}_{\rm{v}}^I$ and overtone at
$\bar{\omega}_{\rm{v}}^{II}$, respectively~\cite{przypis}.
Figs.~\ref{fig:spec_decomp}c,~\ref{fig:spec_decomp}d show isolated
recoil and vibrational contributions of expression \eqref{eq:fit} to
the spectrum of Fig.~\ref{fig:spec_decomp}b. In principle, in
addition to the discussed RIR and vibrational transitions, elastic
transitions with $\delta=0$ are possible both in optical lattice and
for free atoms. However, as shown in Refs.~\cite{guo93,guo94,guo95},
the resulting Rayleigh elastic scattering contribution is negligibly
small in our geometry and polarization configuration.

\section{III. Experiment}

\begin{figure}[h]
\includegraphics[scale=0.78]{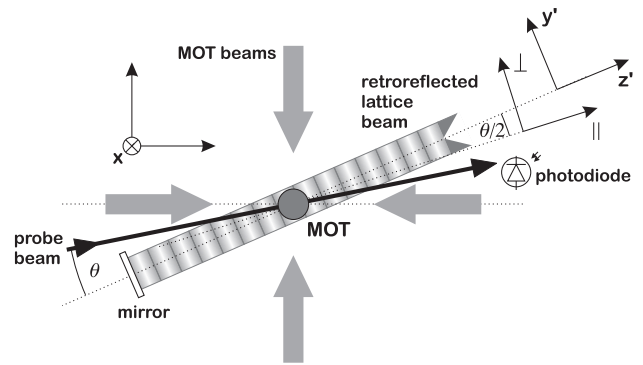}
\caption{\label{fig:setup} Layout of the experiment. The pump and
retroreflected lattice beams intersect in a cloud of cold atoms in
the working MOT. The momentum exchange due to Raman transitions
involving the probe and the lattice photons is allowed in two
directions: $\perp$ and ${||}$. The 1D lattice constrains atomic
motion along the $z$ direction, which, for small $\theta$, nearly
coincides with the ${||}$ direction. The third pair of the MOT beams
and the MOT coils are not shown.}
\end{figure}

In our experiment (see Fig.~\ref{fig:setup}) we use the $^{85}$Rb
atoms trapped and cooled in a standard six-beam, vapor-loaded
MOT~\cite{raab87}. Additional beams intersect at a small angle in
the trap center: the lattice standing wave and the running probe
beam. Both have the same, linear polarization and are blue-detuned
from the trapping transition
$5^2$S$_{1/2}(F=3)$--$5^2$P$_{3/2}(F'=4)$ by
$\Delta_{\rm{latt}}=2\pi\cdot$140 MHz $\approx 23.3\,\Gamma$, where
$\Gamma$ denotes the natural linewidth. The same polarization of the
probe and lattice beams allows a multilevel structure of $^{85}$Rb
to be treated as a set of independent two-level systems. The
relatively big detuning reduces resonant interaction with the atoms;
the scattering rate did not exceed 44 kHz $\approx
7\cdot10^{-3}\,\Gamma/2\pi$ for the lattice beam intensities
$I_{\rm{latt}}=5-35$~mW/cm$^2$. We have measured that such lattice
intensities do not cause any visible heating. The trapping, probe
and lattice beams are generated by diode lasers injection-locked to
a common, external-cavity master oscillator. The probe beam is
scanned by $\delta/2\pi\approx\pm 1$~MHz around frequency $\omega$
of the lattice beam and its transmission is monitored by the
photodiode.

\section{IV. Results}

As already discussed in Section II, in our geometry of the probe and
lattice beams three contributions are expected in the probe
transmission spectrum. Two of them are recoil-induced resonances:
the narrow one, involving the probe and the nearly co-propagating
lattice beam with $\Delta \vc{p}$ along the $\hat{\vc{e}}_\perp$
direction, and the wide one, for the probe and lattice beams nearly
counter-propagating, with $\Delta \vc{p}$ along the
$\hat{\vc{e}}_{||}$ direction (Fig.~\ref{fig:setup}). The third
 contribution is the vibrational spectrum resulting from
Raman transitions between quantized energy levels of the atoms
trapped and oscillating in the 1D lattice
micropotentials~\cite{ver92}. The spectra acquired in our experiment
indeed exhibit all expected contributions, as shown in
Figs.~\ref{fig:spec_decomp}b-d. It proves coexistence of two atomic
fractions in a MOT: the free moving atoms and the atoms captured by
the 1D lattice. Thanks to the large detuning $\Delta_{\rm{latt}}$,
resonances induced by the lattice and probe beams are not influenced
by those involving the MOT-beams. This is a significant
simplification with respect to the previous work~\cite{brzo051},
allowing spectroscopic discerning of two stable atomic fractions and
sensitive study of their coexistence.

\begin{figure}[h]
\includegraphics[scale=0.66]{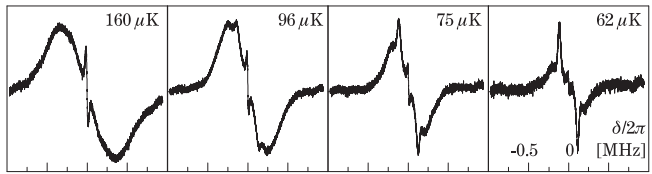}
\caption{\label{fig:lattice_emergence}Comparison of the probe
transmission spectra for various MOT-beam temperatures. For
sufficiently low temperatures the wide RIR contribution disappears
while the lattice contribution becomes distinct. The sequence of
spectra illustrates well the competition between the free and bound
atoms. Vertical scale is the same for all plots.}
\end{figure}

Fig.~\ref{fig:lattice_emergence} presents the series of measurements
taken for different temperatures of the atomic cloud. The
temperature is controlled by the MOT beam intensity and is
determined by the width of the wide RIR. At temperatures about the
Doppler limit, $T_D=140\,\mu$K, atoms localized in lattice wells
become a significant fraction of the whole sample and give rise to a
pronounced vibrational structure.

\begin{figure}[h]
\includegraphics[scale=0.5]{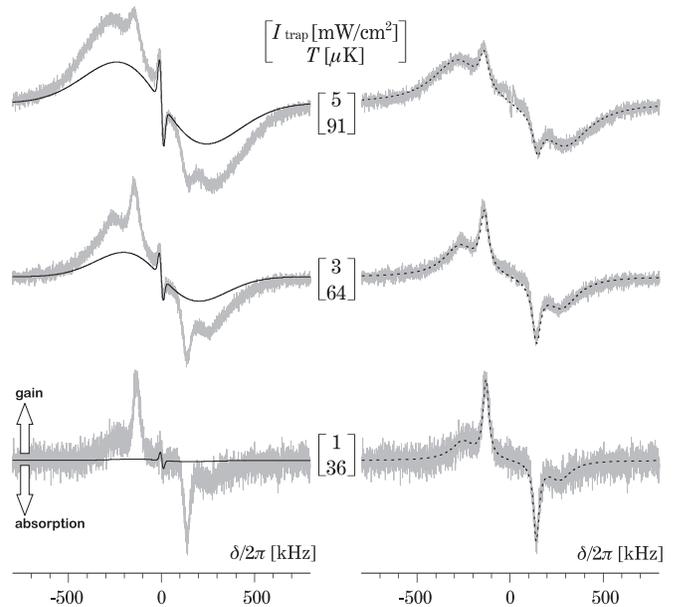}
\caption{\label{fig:probe_trans} Probe transmission spectra with 1D optical
lattice recorded for various MOT-beam intensities and related temperatures.
Left column, measured spectra (grey) and RIR contributions (solid black)
determined from two-beam thermometry (see text). Right column, vibrational
contributions extracted by substraction of RIR components from full spectra and
theoretical fits (dotted). Vertical scale is the same for all plots.}
\end{figure}

In the spectra measured with three-beam configuration the first
vibrational component is very distinct but the overtone overlaps
with the wide RIR, which greatly obstructs interpretation of the
lattice spectra. This obstacle can be removed by a thorough
separation of RIR and lattice contributions. For this sake, we
reduced the number of free parameters by determining the temperature
of the cold atomic cloud for each measurement. This was done by
blocking the mirror that retroreflects the lattice beam
(Fig.~\ref{fig:setup}) and by performing two-beam RIR velocimetry
with nearly counter-propagating Raman beams~\cite{brzo052,mea94}. In
such a configuration, the only contribution to the spectrum is the
RIR associated with the momentum transfer along the
$\hat{\vc{e}}_{||}$ direction (Fig.~\ref{fig:spec_decomp}a). Since
our 1D optical lattice does not exhibit spatial polarization
modulation, no additional cooling mechanism is present. This allows
us to assume that the temperature of atoms in a MOT with and without
the lattice is the same. Having determined the width $\xi_{||}$, we
calculated the temperature $T$ and the width
$\xi_{\perp}=\xi_{||}\tan(\theta/2)$~\cite{brzo051,brzo052}. Next,
the widths $\xi_{||}$ and $\xi_{\perp}$ were used as constants in
the first terms of the fit given by eq.~\eqref{eq:fit} and then the
fitted RIR contributions were subtracted from the spectra recorded
with both lattice beams. The resulting difference constitutes pure
lattice contribution which is very well reproduced by the last terms
of eq.~\eqref{eq:fit}, as seen in Fig.~\ref{fig:probe_trans}.

For decreasing temperature of the atomic cloud, we observed increase
of the amplitude of the vibrational contribution relative to the RIR
contribution. This indicates a growth of the fraction localized in
the 1D optical lattice. We also observed systematic variation of the
relative amplitudes of the two RIR contributions. As it can be seen
in Fig.~\ref{fig:probe_trans}, for lower temperatures the amplitude
of the wide RIR associated with the momentum exchange in the $||$
direction is noticeably smaller than the amplitude of the narrow RIR
determined by the exchange in the $\perp$ direction. This behavior
can be explained by constraints imposed by the 1D optical lattice on
the atomic movement resulting in suppression of the recoil in the
$||$ direction. Hence, localized atoms do not contribute to the wide
RIR.  Since the increase of the lattice-trapped fraction caused by
the temperature lowering proceeds at the expense of unbound atoms,
the amplitude of the wide RIR becomes smaller. Inversely, since
there is no lattice trapping in the $\perp$ direction, the
localization does not affect the narrow RIR, the amplitude of which
depends only on the total number of atoms. Further lowering of the
temperature leads to the complete suppression of the wide RIR, which
indicates that all atoms still remaining in a MOT are loaded into 1D
optical lattice. This qualitative consideration suggest the way of
controlling the ratio of the lattice-localized to unbound atoms.

\begin{figure}[h]
\includegraphics[scale=0.6]{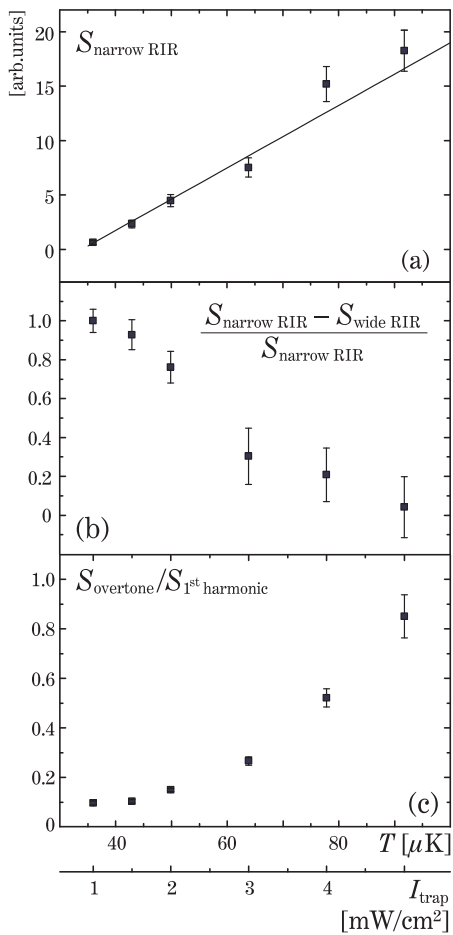}
\caption{\label{fig:plotfields} Trap-beam intensity and temperature
dependence of the atomic fractions estimated by calculating areas
$S$ associated with relevant spectra contributions: (a)
$S_{\rm{narrow\,\,RIR}}$, re-scaled as described in text,
proportional to the total number of atoms in a MOT, (b)
$(S_{\rm{narrow\,\,RIR}}-S_{\rm{wide\,\,RIR}})/S_{\rm{narrow\,\,RIR}}$,
the atomic fraction localized in a 1D optical lattice, (c)
$S_{\rm{overtone}}/S_{\rm{1^{\rm{st}}\,\,harmonic}}$, the ratio of
atoms undergoing Raman transitions responsible for the 1$^{\rm{st}}$
harmonic and overtone.}
\end{figure}

To examine the behavior of the discussed atomic fractions
quantitatively, we numerically estimated the number of atoms which
give rise to the relevant terms of eq.~\eqref{eq:fit}. Since RIR
signals are proportional to derivative of atomic momentum
distributions, we first reconstructed their integrals and then
evaluated the areas under resulting curves. In case of lattice
contributions we calculated the areas directly under the vibrational
Raman signals.

Fig.~\ref{fig:plotfields}a presents the area
$S_{\rm{narrow\,\,RIR}}$ as a function of MOT temperature.
$S_{\rm{narrow\,\,RIR}}$ is obtained by calculating the area under
momentum distribution derived from the first term of
eq.~\eqref{eq:fit} and dividing by the geometrical factor
$\tan^2(\theta/2)$~\cite{brzo051,brzo052} for comparison with
$S_{\rm{wide\,\,RIR}}$, the area associated with the second term of
Eq.~\eqref{eq:fit}. Since there is no lattice trapping in this
direction, $S_{\rm{narrow\,\,RIR}}$ is proportional to the total
number of atoms in the trap. In consistency with previous
observations~\cite{wieman92} it decreases linearly when MOT-beam
intensity is lowered.

Fig.~\ref{fig:plotfields}b depicts the ratio
$(S_{\rm{narrow\,\,RIR}}$--$S_{\rm{wide\,\,RIR}})$/
$S_{\rm{narrow\,\,RIR}}$ which, due to the same transition
amplitudes for both RIRs, is a direct measure of the fraction
captured by 1D optical lattice. As expected, the localized fraction
grows with decreasing temperature. For the temperature of 30~$\mu$K
all the MOT-trapped atoms are in the optical lattice. The absolute
number of atoms localized in the lattice also begins to grow with
decreasing temperature, but then the general decrease of the total
number of atoms in the MOT prevails. The number of atoms in the
lattice reaches its maximum at about 60~$\mu$K when the captured
fraction is close to 0.5.

For growing temperatures we observed a systematic increase of the
overtone contribution relative to the $1^{\rm{st}}$ harmonic
($S_{\rm{overtone}}/S_{\rm{1^{\rm{st}}\,\,harmonic}}$ ratio,
depicted in Fig.~\ref{fig:plotfields}c). The rising probability of
the overtone transitions is attributed to the fact that for higher
temperature population is shifted towards higher vibrational levels
with more pronounced anharmonicity~\cite{herzberg}.

\begin{figure}[ht]
\includegraphics[scale=0.52]{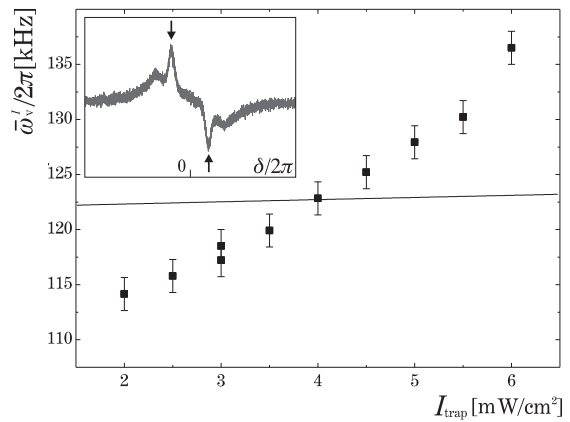}
\caption{\label{fig:omegavib} Frequency
$\bar{\omega}_{\rm{v}}^{\,I}$ of the vibrational resonance (marked
by arrows on the lattice spectrum in the inset) as a function of the
MOT trapping beams intensity. The solid line represents predictions
of independent light-shifts, as described in text.}
\end{figure}

We have also studied changes of the position of the average
first-harmonic frequency, $\bar{\omega}_{\rm{v}}^{\,I}$, with the
increase of the trapping-beam intensity, proportional do atomic
temperature (Fig.~\ref{fig:omegavib}). Taking only the temperature
effect into account, one might expect a shift of the recorded
vibrational resonances towards lower frequencies with the increasing
temperature, since the higher vibrational levels become more
populated.
 However, the observed dependence is opposite:
$\bar{\omega}_{\rm{v}}^{\,I}$ increases when the trapping light
becomes more intense. We interpret this effect by including also the
 light shift effect. The
levels of the atoms in a MOT with the lattice beams switched on are
perturbed by two fields: the red-detuned MOT field and the
blue-detuned lattice field. When the intensity of the trapping light
is increased, the related light-shifts of the ground atomic
sublevels also increase. This decreases the detuning of the lattice
beams from the atomic resonance frequency. Since the depth of the
lattice potential $U_{\rm{latt}}$ is proportional to $I_{\rm{latt}}
/ \Delta_{\rm{latt}}$, we observe the increase of $U_{\rm{latt}}$
followed by increase of $\bar{\omega}_{\rm{v}}^{\,I}\propto
\sqrt{U_{\rm{latt}}}$. The calculations based on the assumption of
independent light shifts associated with the MOT and lattice beams
(solid line in Fig.~\ref{fig:omegavib}), similar to those performed
for far-detuned light~\cite{katori}, predict the correct slope, but
fail to reproduce the size of the observed dependence of
$\bar{\omega}_{\rm{v}}$ which indicates need of more refined
calculations.

\section{V. Conclusions}

We have performed pump-probe spectroscopy of $^{85}$Rb atoms in an
operating MOT equipped with an extra pair of lattice beams,
blue-detuned by more than $20\,\Gamma$ from the trapping transition
$F=3$ -- $F'=4$. Application of a simple 1D lattice of a
standing-wave configuration and nondestructive RIR thermometry
allowed studies of the localization efficiency at well-determined
atomic temperature, not affected by the lattice. The recorded
pump-probe spectra revealed complex structure composed of the
recoil-induced and vibrational resonances due to the free, and
lattice-bound atoms, respectively. The magnitude of each of these
spectral features yielded information on the population of the
corresponding atomic fraction. We have shown how to resolve the
pump-probe spectra for an atomic system consisting of two atomic
fractions and applied this technique to the spectra taken at various
atomic temperatures. This extension of a standard RIR velocimetry
proved very useful for nondestructive diagnostics of 1D-lattice
loading. We were able to verify that the creation of a robust 1D
optical lattice is possible by loading of relatively hot
($\sim$100~$\mu$K) atoms directly from a continuously operating
regular MOT without any further cooling.

\subsection{Acknowlegements}
This work is supported by the Polish Ministry of Science and
Information Society Technologies and is part of a general program on
cold-atom physics of the National Laboratory of AMO Physics in
Toruń, Poland. We would like to thank Julien Saby for his assistance
in analysis of the experimental data and Dmitry Budker and Krzysztof
Sacha for their remarks on the manuscript.


\begin{thebibliography}{99}

\bibitem{gryrob01} for review see, for example, G. Grynberg and C. Robilliard, Phys. Rep. \textbf{355}, 335
(2001) and references therein.
\bibitem{dali89} J. Dalibard and C. Cohen-Tannoudji, J. Opt. Soc. Am. B \textbf{6}, 2023
(1998).
\bibitem{lett88} P.D. Lett, R.N. Watts, C.I. Westbrook, W.D. Phillips, P.L. Gould and H.J.
Metcalf, Phys. Rev. Lett. \textbf{61}, 169 (1988).
\bibitem{greiner01} see, for example, M. Greiner, I. Bloch, O. Mandel, T.W. Hänsch and T. Esslinger, Appl.
Phys. B \textbf{73}, 769-772 (2001) and references therein.
\bibitem{petsas79} K.I. Petsas, A.B. Coates and G. Grynberg, Phys. Rev. A, \textbf{50}, 5173 (1994).
\bibitem{schad99} H. Schadwinkel, U. Reiter, V. Gomer and D. Meschede, Phys. Rev. A, \textbf{61} 013409 (1999).
\bibitem{ver92} P. Verkerk, B. Lounis, C. Salomon, and C. Cohen–Tannoudji, J.–Y. Courtois and G. Grynberg, Phys. Rev. Lett. \textbf{68}, 3861, (1992).
\bibitem{hemm93} A. Hemmerich and T.W. H\"{a}nsch, Phys. Rev. Lett. \textbf{70}, 410 (1993).
\bibitem{jessen92} P.S. Jessen, C. Gerz, P.D. Lett, W.D. Phillips, S.L. Rolston, R.J.C. Spreeuw, and C.I. Westbrook, Phys. Rev. Lett. \textbf{69} 49 (1992)
\bibitem{kozuma95} M. Kozuma, Y. Imai, K. Nakagawa, and M. Ohtsu, Phys. Rev. A \textbf{52}, R3421 (1995).
\bibitem{kozuma96} M. Kozuma, K. Nakagawa, W. Jhe, and M. Ohtsu, Phys. Rev. Lett. \textbf{76}, 2428 (1996).
\bibitem{guo93} J. Guo and P. Berman, Phys. Rev. A \textbf{47}, 4128 (1993).
\bibitem{guo92} J. Guo, P. R. Berman and B. Dubetsky, Phys. Rev. A \textbf{46}, 1426 (1992).
\bibitem{guo94} J. Guo, Phys. Rev. A \textbf{49}, 3934 (1994).
\bibitem{guo95} J. Guo, Phys. Rev. A \textbf{52}, 1458 (1995).
\bibitem{brzo052} M. Brzozowska, T.M. Brzozowski, J. Zachorowski and W. Gawlik, \textbf{72}, 061401(R) (2005).
\bibitem{gry94} G. Grynberg. J.-Y. Courtois, B. Lounis, and P. Verkerk, Phys. Rev. Lett. \textbf{72}, 3017 (1994).
\bibitem{ver96} P. Verkerk, Proceedings of the International School of Physics, Varenna, Course \textbf{CXXXI}, 325 (1996).
\bibitem{raab87} E. L. Raab, M. Prentiss, A. Cable, S. Chu, and D. E. Pritchard, Phys. Rev. Lett. \textbf{59}, 2631 (1987).
\bibitem{lett92} P.S. Jessen, C. Gerz, P.D. Lett, W.D. Phillips, S.L. Rolston, R.J.C. Spreeuw, and C.I.
Westbrook, Phys. Rev. Lett., \textbf{69} 49 (1992).
\bibitem{brzo051} T. M. Brzozowski, M. Brzozowska, J. Zachorowski, M. Zawada, and W. Gawlik, Phys. Rev. A \textbf{71}, 013401
(2005).
\bibitem{przypis} Our calculations based on C. Cohen-Tannoudji, J. Dupont-Roc and G. Grynberg, \textit{Atom-Photon Interactions: Basic Processes and
Applications}, John Wiley {\&} Sons, 1992, pp. 86-87; 519-523;
proved that higher overtones can be neglected.
\bibitem{mea94}D. R. Meacher, D. Boiron, H. Metcalf, C. Salomon, and G. Grynberg, Phys. Rev. A \textbf{50},  R1992 (1994).
\bibitem{wieman92} see, for example, K. Lindquist, M. Stephens,
and C. Wieman, Phys. Rev. A \textbf{46}, 4082 (1992).
\bibitem{herzberg} G. Herzberg,  \textit{Molecular Spectra and Molecular Structure}, Krieger Publishing
Company 1992, pp.85-88
\bibitem{ficek93} e.g. Z. Ficek and H.S. Freedhoff, Phys. Rev. A, \textbf{48} 3092
(1993).
\bibitem{katori} see, for example, T. Ido and H. Katori, Phys. Rev. Lett, \textbf{91}
053001 (2003).
\end{thebibliography}
\end{document}